\documentclass{PoS}
\usepackage{graphicx}
\usepackage{placeins}

\newcommand{\pt}{\ensuremath{p_{\mathrm{T}}}}

\newcommand{\ptgamma}{\ensuremath{p_{\mathrm{T}^\gamma}}}
\newcommand{\pth}{\ensuremath{p_{\mathrm{T}^h}}}

\newcommand{\pPb}{p--Pb}
\newcommand{\PbPb}{Pb--Pb}

\newcommand{\zt}{\ensuremath{z_{\mathrm{T}}}}

\newcommand{\GeVc}{\ensuremath{\mathrm{GeV}/c}}
\newcommand{\sqrts}{\ensuremath{\sqrt{s}}}

\newcommand{\ydecay}{\ensuremath{\gamma^\mathrm{decay}}}
\newcommand{\gammaiso}{\ensuremath{\gamma^\mathrm{iso}}}

\newcommand{\sqrtsNN}{\ensuremath{\sqrt{s_\mathrm{NN}}}}

\usepackage{lineno}

\title{Measurement of isolated photon-hadron and jet correlations in 5 TeV pp and \pPb~collisions with the ALICE detector at the LHC}

\ShortTitle{Isolated photon-hadron and jet correlations in pp and p--Pb collisions}

\author{\speaker{Miguel Arratia}, on behalf of the ALICE collaboration\\
        University of California, Berkeley.\\
        E-mail: \email{marratia@berkeley.edu}}

\abstract{Photon-jet correlations are a promising channel for the study of parton energy loss in nuclear collisions. While existing measurements in pp and nuclear collisions have used high energy photons and jets, we focus on an unexplored kinematic range given by $12<p_{\mathrm{T}}<30$ GeV/$c$ photons and the corresponding low jet $p_{\mathrm{T}}$. We present results obtained using 5.02 TeV pp and \pPb~collisions. A combination of isolation and electromagnetic shower-shape variables is used to reduce the large background from meson decays and fragmentation photons. We show how the access to this kinematic range of hard probes was achieved with a novel combination of high rate and low-momentum tracking using the electromagnetic calorimeters and the inner tracking system of the ALICE experiment.}

\FullConference{International Conference on Hard and Electromagnetic Probes of High-Energy Nuclear Collisions\\
		30 September - 5 October 2018\\
		Aix-Les-Bains, Savoie, France}

\begin{document}

\section{Introduction}
Photon-jet correlations are a promising channel for the study of parton energy loss in nuclear collisions.
 The comparison of pp, \pPb, and \PbPb~data disentangles effects due to the quark--gluon plasma and effects such as modification of parton distribution functions in nuclei and multiple parton scattering processes inside a large nucleus. This is because final-state effects associated with the quark--gluon plasma are expected to be absent or suppressed in pp and \pPb~collisions.

\section{Experimental setup and data sets}
A comprehensive description of the ALICE experiment and its performance is provided in Ref.~\cite{Abelev:2014ffa}. The detector elements most relevant for this study are the ElectroMagnetic Calorimeter (EMCal) and the Inner Tracking System (ITS); both are within a 0.5 T solenoidal magnetic field. 

The EMCal is a lead-scintillator sampling calorimeter with towers arranged in a quasi-projective geometry. Its granularity in pseudorapidity and azimuthal angle is $\Delta\eta\times\Delta\varphi$ = 14.3$\times$14.3 mrad$^{2}$, and its energy resolution is $\sigma_{E}/E = 4.8\%/E\otimes 11.3\%/\sqrt{E}\otimes 1.7\%$, with the energy $E$ in units of GeV~\cite{Abeysekara:2010ze}. Its acceptance is $|\eta|<0.70$ and 1.396 $< \varphi <$ 3.264 rad.

The ITS consists of six layers of silicon detectors and is located around the interaction point covering 3.9 $<r<$ 43 cm, $|\eta|<0.9$, and $0< \varphi<2\pi$ rad.

The analyzed  data were collected during the \sqrtsNN{} = 5 TeV \pPb~run in 2013 and during the \sqrts{} = 5 TeV pp run in 2017. 
One of the novel aspects of this analysis is the use of ITS standalone tracking (excluding time projection chamber), which was exploited to increase the data taking rate with partial readout. The total integrated luminosity is about {4.6 nb$^{-1}$} and about {300 nb$^{-1}$} for the 2013 \pPb~and the 2017 pp data sample, respectively. 

\subsection{Photon reconstruction}
In this analysis, the signal are ``prompt" photons, which include ``direct photons" and ``fragmentation photons''. At leading order in perturbative QCD, the direct photons are produced in hard scattering processes such as gluon Compton scattering ($qg\to q\gamma$) or quark-antiquark annihilation ($q\bar{q}\to g\gamma$), whereas the fragmentation photons are the product of the collinear fragmentation of a parton ($q\bar{q}(gg)\to \gamma + X$). 

The fragmentation contribution to the total cross section can be reduced using an isolation criterion, which also suppresses the background from decays of neutral mesons. The isolation variable is defined as the scalar sum of the transverse momentum of charged particles within an angular radius around the cluster direction, $R =\sqrt{(\Delta\varphi)^{2} +(\Delta\eta)^{2}  } =0.4$. The underlying event is subtracted by using the charged-particle density, $\rho$, obtained with the \textsc{FastJet} jet area/median method~\cite{Cacciari:2009dp}:
\begin{equation}
ISO = \sum_{\mathrm{track}~\in\Delta R<0.4} p_{\mathrm{T}}^{\mathrm{track}} - \rho \times \pi(0.4)^{2}.
\label{eq:isoraw}
\end{equation}

A selection of $ISO<1.5$ GeV/$c$ reduces the fragmentation contribution to below 20$\%$ according to \textsc{JETPHOX} calculations~\cite{JETPHOX}. The main background of this selection arises from multi-jet events where one jet typically contains a $\pi^{0}$ or $\eta$ that carries most of the jet energy and is misidentified as a photon.

The measurement of the signal purity of the isolated photon selection is performed with the ``template-fit" method, in which the measured shower-shape distribution is fit with the sum of a signal and background templates with the relative normalization as the single free parameter. The background template is obtained with an isolation sideband, while the signal template is obtained from simulation. This is a new development for photon reconstruction in ALICE. 

\begin{figure}[h]
\center
\includegraphics[width=.53\textwidth]{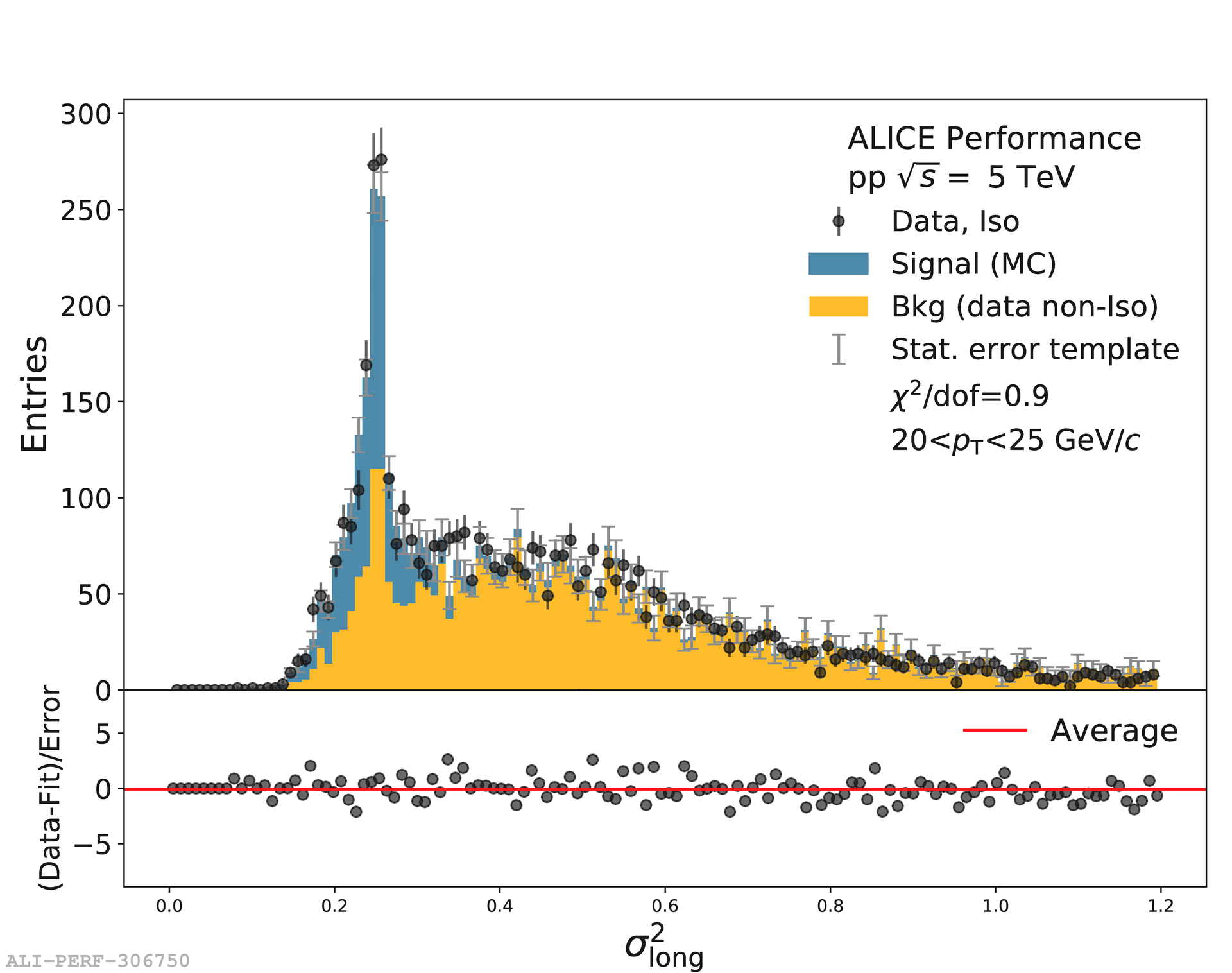}
\includegraphics[width=.42\textwidth]{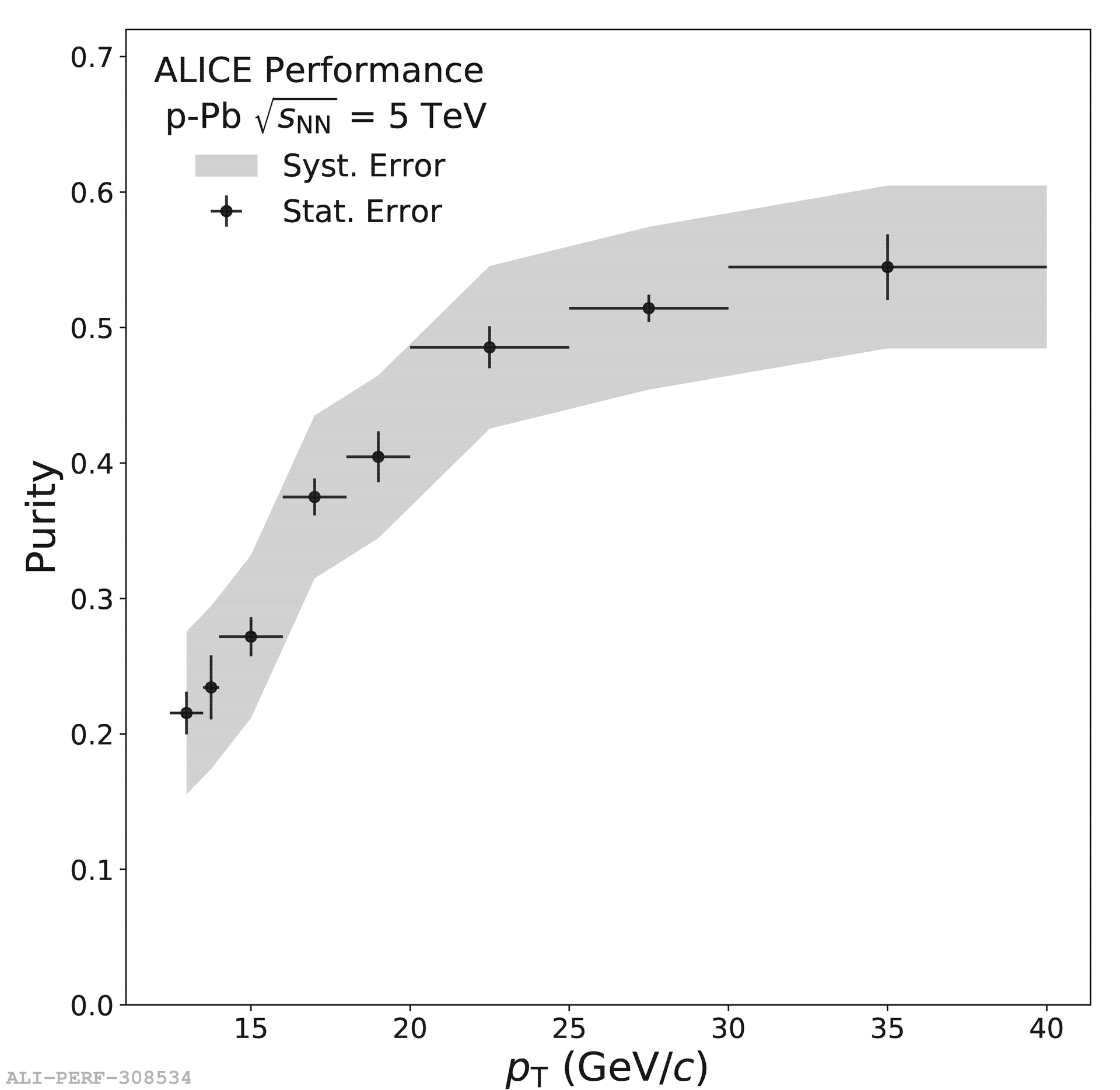}
\caption{Left: Template fit results in pp collisions using the $\sigma_{\mathrm{long}}^{2}$ variable. Right: Purity of isolated photon selection as a function of $p_{\mathrm{T}}$.}
\label{SidebandTemplateFit}
\end{figure}

Figure~\ref{SidebandTemplateFit} shows an example of the template fit with the $\sigma_{\mathrm{long}}^{2}$ variable that is a energy-weighted RMS of the shower-shape profile~\cite{Abeysekara:2010ze}, and the resulting purity of the isolated photon , $\gammaiso$, selection: $\sigma_{\mathrm{long}}^{2}<0.3$ and $ISO<1.5$ GeV/$c$. 

\section{Isolated photon-hadron correlations}
\label{GammaHadron}
To obtain the angular correlation between $\gammaiso$ and hadrons, several corrections are applied: geometrical acceptance effects are corrected by using the mixed-event technique; the uncorrelated background is estimated by using a control region at large $|\eta^{\mathrm{hadron}}-\eta^{\gamma}|$; the magnitude and shape of the correlated 
{$\ydecay$--hadron} background is estimated from the measured purity and from an inversion of the shower-shape selection, respectively. This analysis is performed with photons with $12<\ptgamma{} <15~\GeVc$, and in intervals of $\zt \equiv \pth/\ptgamma$ in the range $0.12 < \zt < 0.42$. 

\begin{figure}[h]
\center
\includegraphics[width=.49\textwidth]{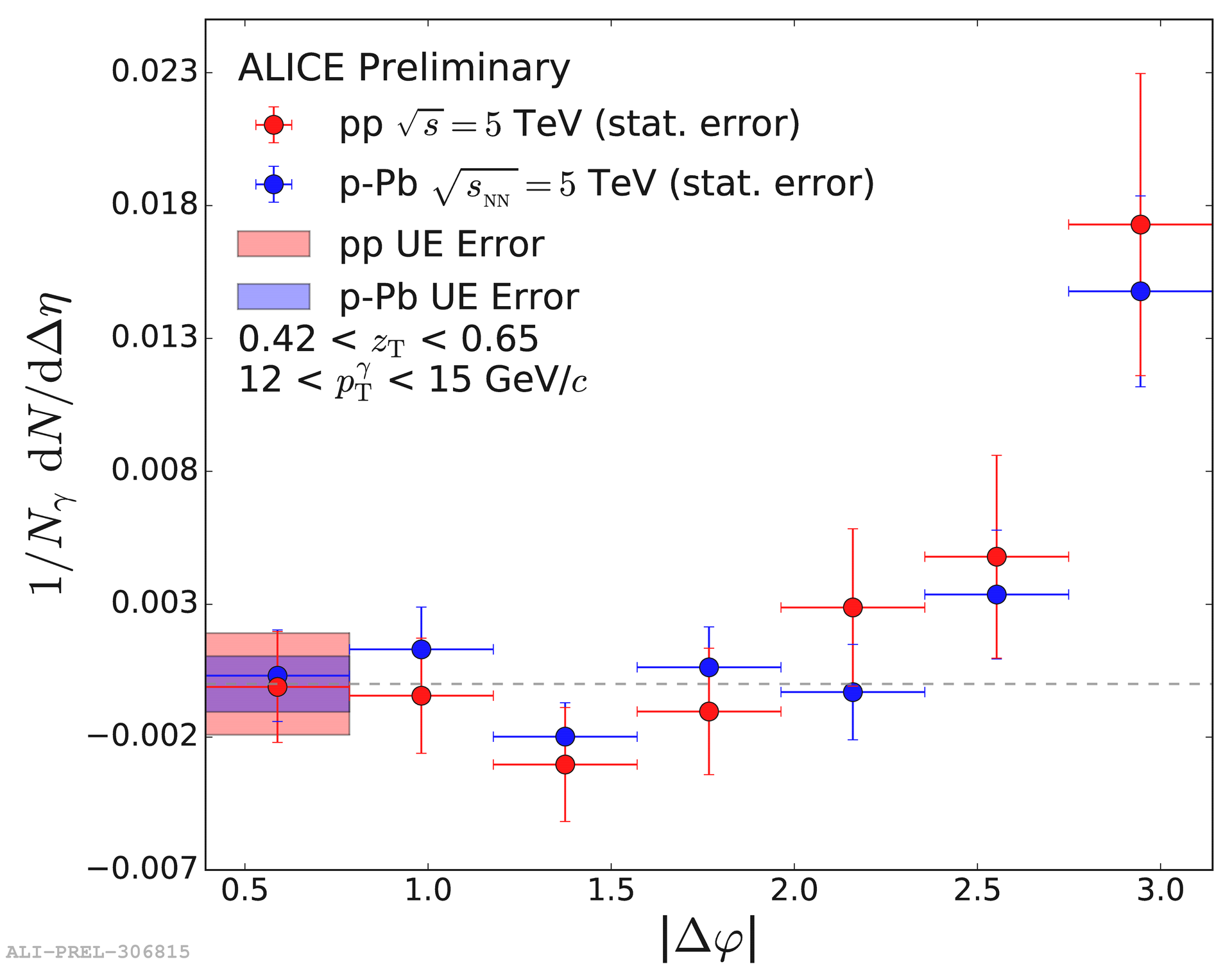}
\includegraphics[width=.49\textwidth]{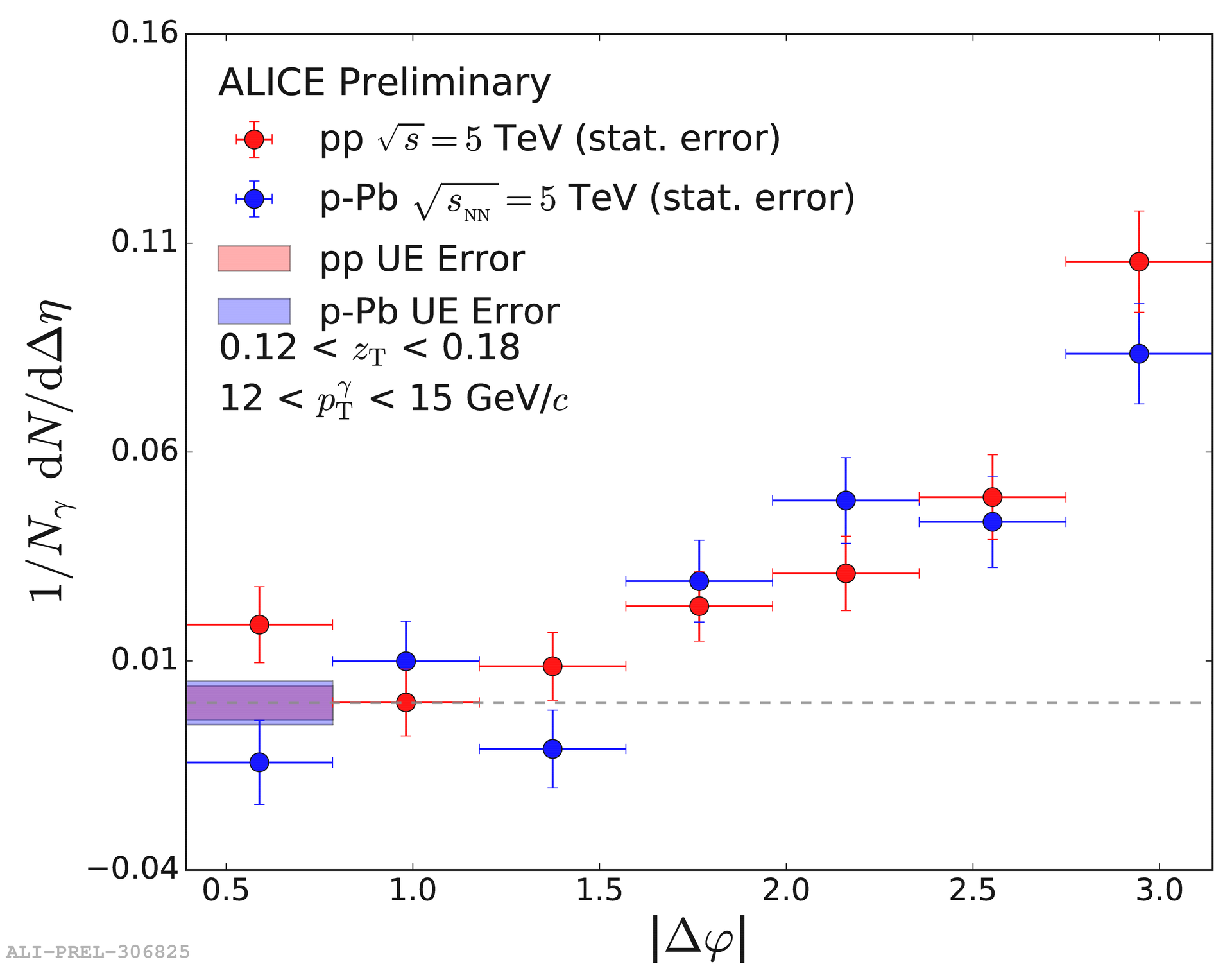}
\caption{\gammaiso--hadron correlations in different \zt~bins.}
\label{FragmentationFunctionZTBINS}
\end{figure}

\begin{figure}[h]
\center
\includegraphics[width=.55\textwidth]{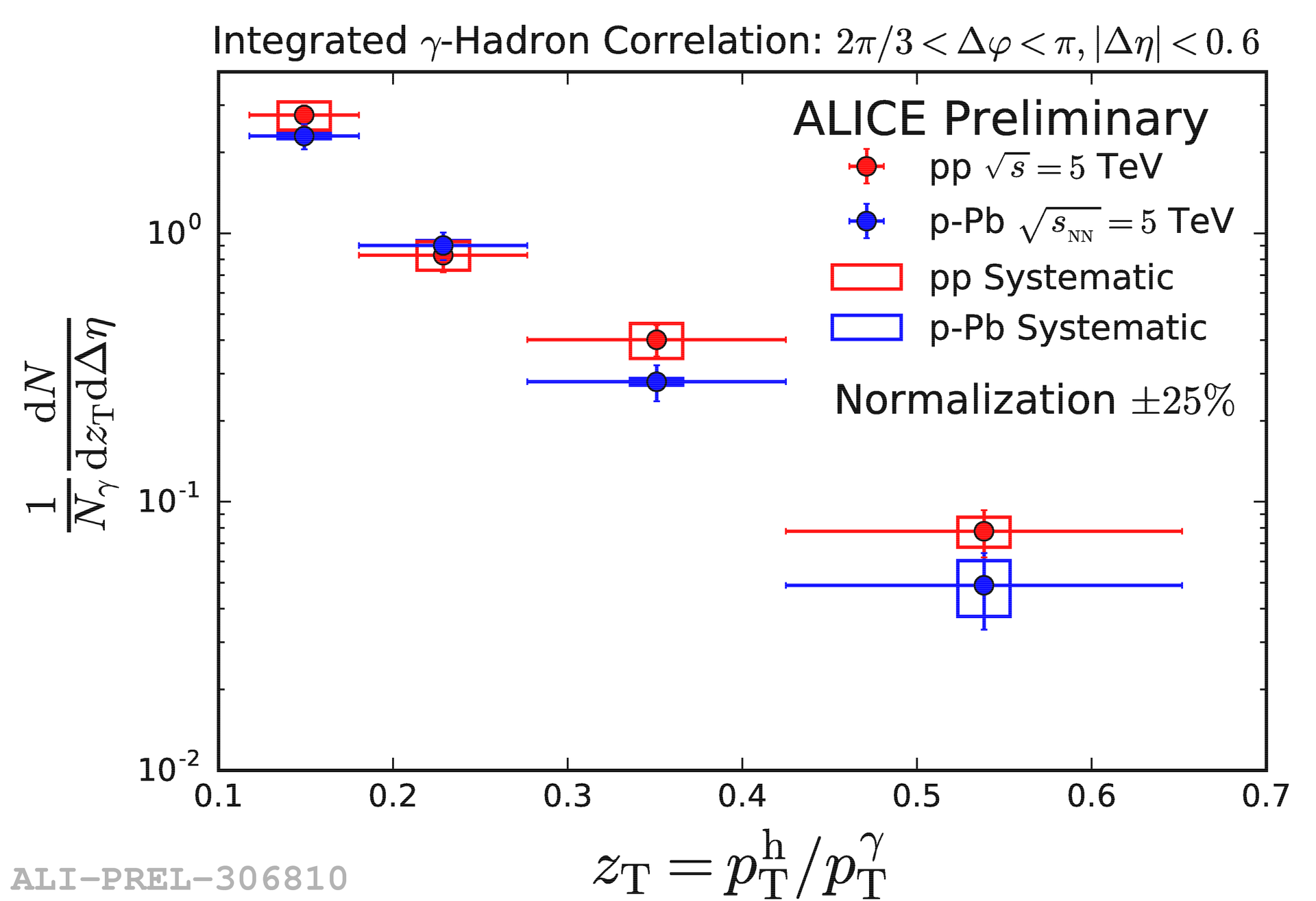}
\caption{Integrated correlations between isolated photons and hadrons as a function of $z_{\mathrm{T}}$. }
\label{FragmentationFunction}
\end{figure}

Figure~\ref{FragmentationFunctionZTBINS} shows the measured \gammaiso--hadron correlations measured in pp and \pPb~collisions. As expected, the signal peaks at $\Delta\varphi = \pi$; the peak gets narrower with increasing $\zt$. The data are compatible within uncertainties.  Figure~\ref{FragmentationFunction} shows the integrated correlation function in the range {$\Delta\varphi> 2\pi/3$} and $\Delta\eta<0.6$. The pp and \pPb~results are compatible within uncertainties. 

\section{Isolated photon-jet correlations}
\begin{figure}[h]
\center
\includegraphics[width=.49\textwidth]{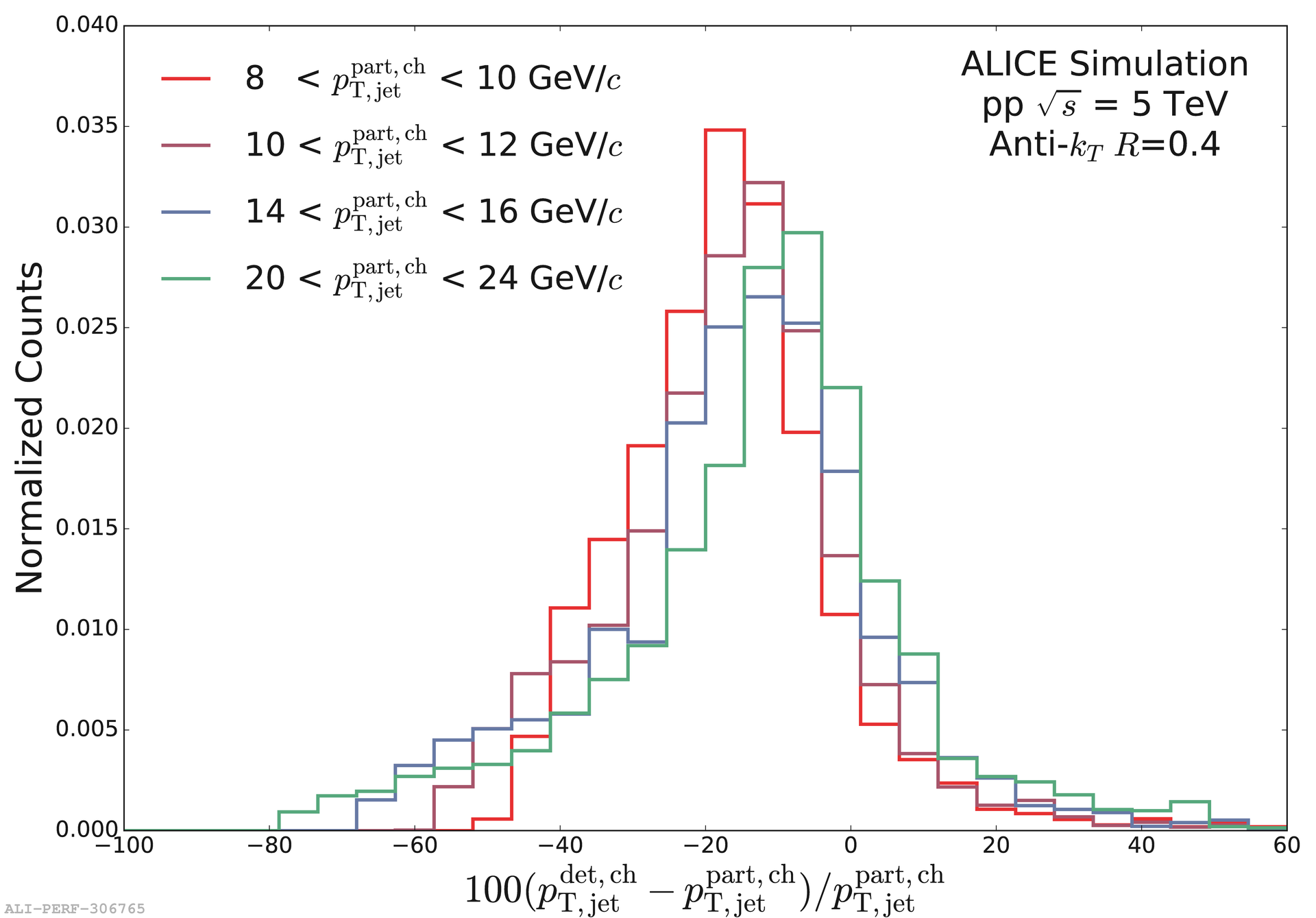}
\includegraphics[width=.49\textwidth]{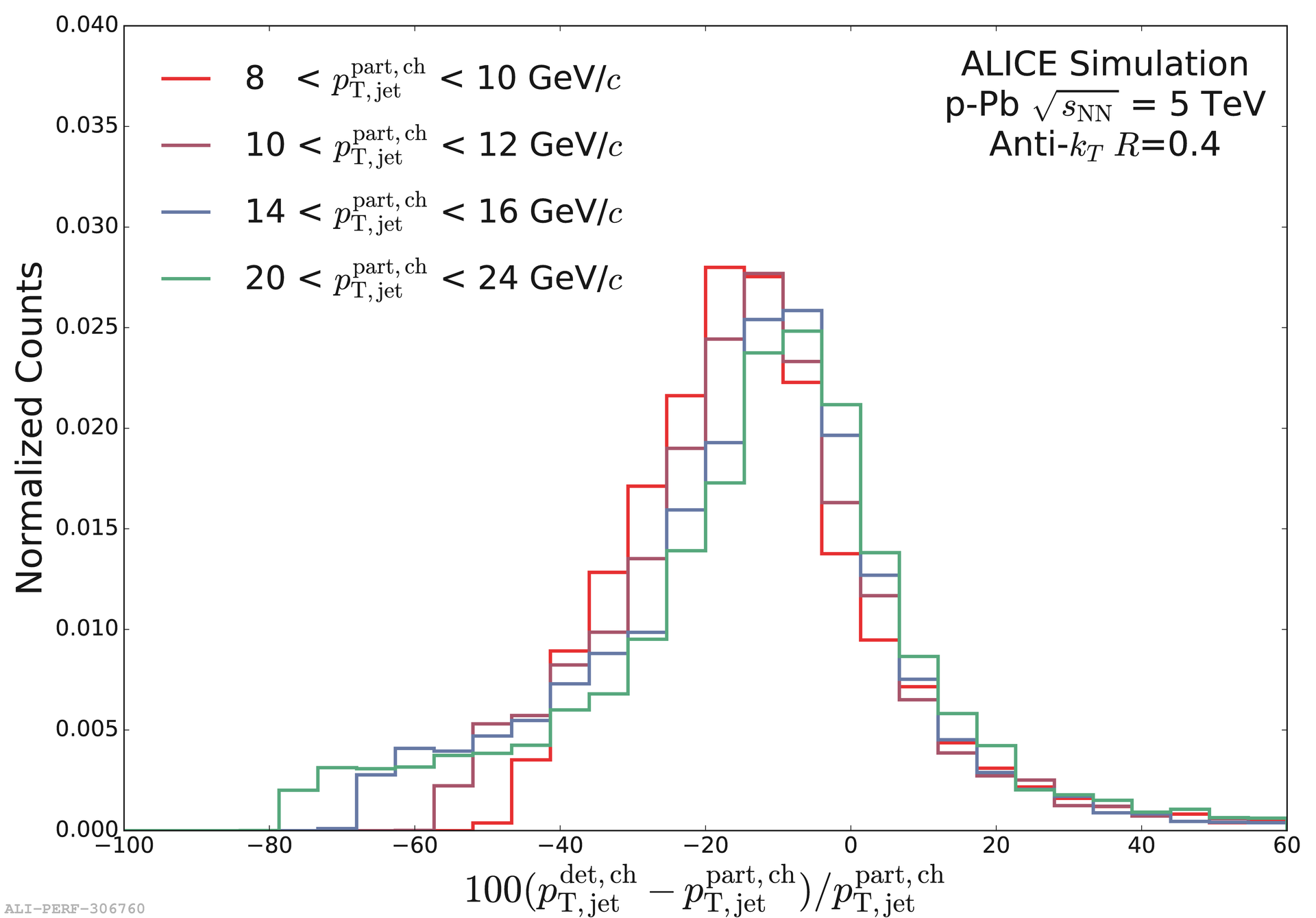}
\caption{Charged-jet transverse momentum resolution for different ranges in true charged-jet transverse momentum, obtained with ITS-only reconstruction.}
\label{Resolution}
\end{figure}

Jets are reconstructed with the anti-$k_{\mathrm{T}}$ algorithm on tracks with {$0.15$ $<\pt < 15$ \GeVc} and $|\eta|<0.8$ as input. Figure~\ref{Resolution} illustrates the jet momentum resolution for pp and \pPb~collisions using ITS-only tracking. No large differences in performance are observed between pp and \pPb~data sets. The bias of about 15$\%$ is due to tracking efficiency and the resolution is driven by the track momentum resolution (which is about 7 times worse than for the standard ALICE tracking).  


The subtraction of the $\ydecay$--jet background arising from the impurity of the $\gammaiso$ selection follows the procedure described in Sec.~\ref{GammaHadron}. The background produced by jets that originate from the underlying event is estimated with an event-mixing approach; this represents about $10\%$ of the jet yield for \pPb~collisions and is negligible for pp collisions.  

\begin{figure}[h]
\center
\includegraphics[width=.49\textwidth]{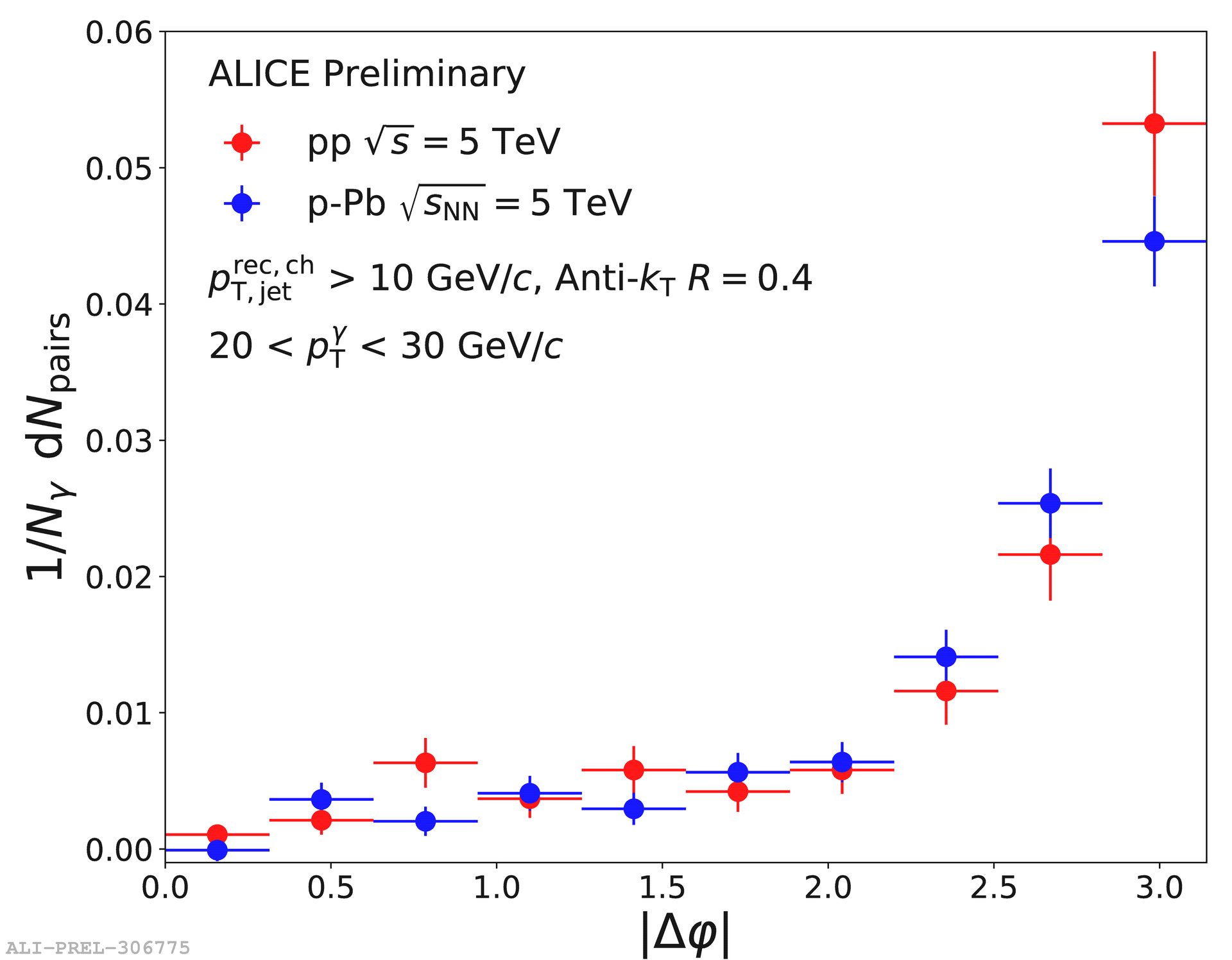}
\includegraphics[width=.49\textwidth]{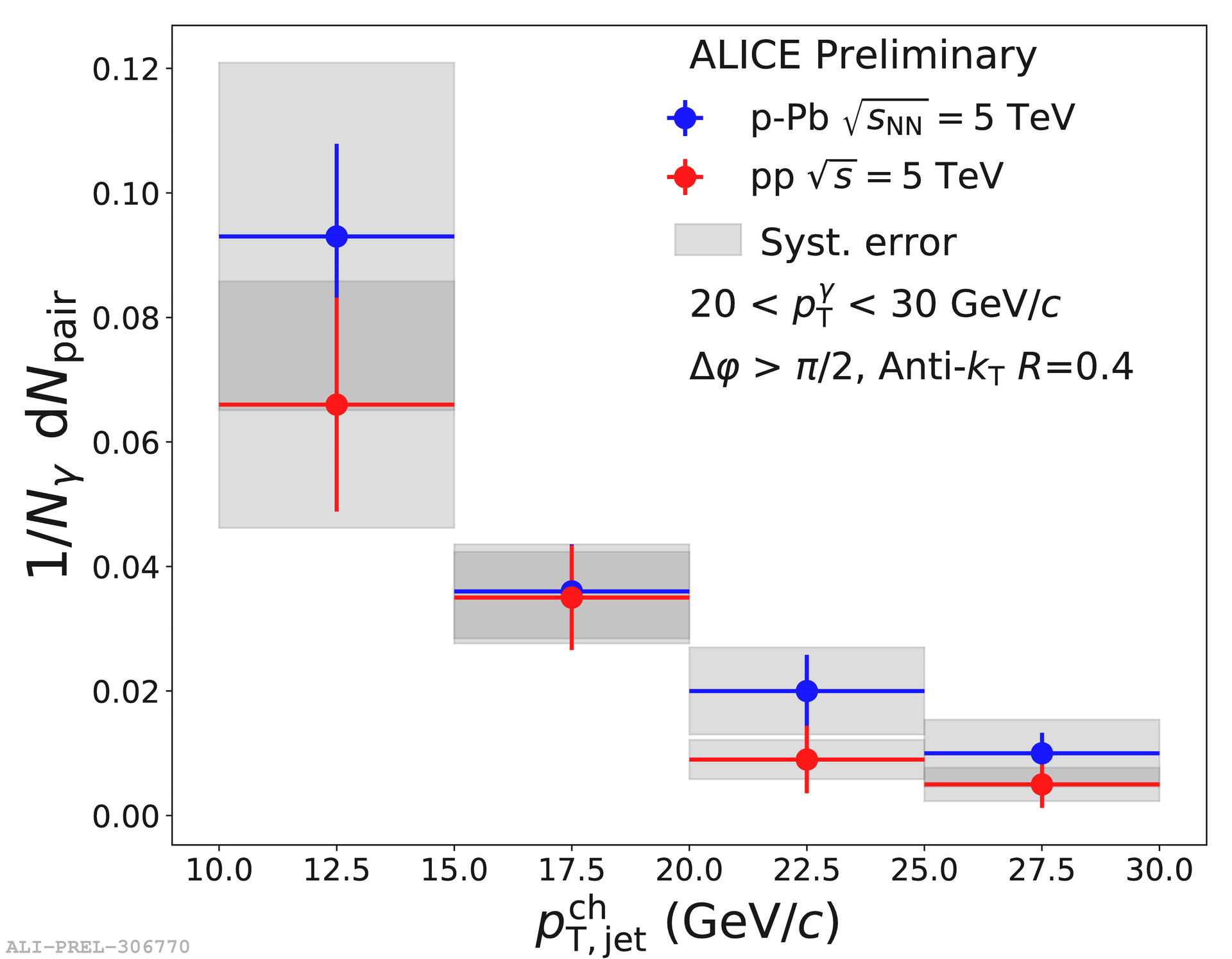}
\caption{Azimuthal difference between isolated photon candidates and charged-jets. Yield of jets per isolated photon as a function of jet transverse momentum. }
\label{GammaJet}
\end{figure}

Figure~\ref{GammaJet} shows the azimuthal correlation between $\gammaiso$ candidates and charged jets, which peaks at $|\Delta\varphi|=\pi$, and the transverse momentum spectrum of recoiling jets. In both cases the pp and \pPb~data are compatible within uncertainties. 

\section{Conclusions}
A measurement of photon-hadron and photon-jet correlations in \pPb~and pp collisions at {5 TeV} is reported. This is the first analysis of its kind by the ALICE collaboration. The results indicate that there is no difference between the two data sets within uncertainties. This result establishes a benchmark on photon identification and jet reconstruction for future ALICE measurements.

\renewcommand\refname{Bibliography}

\end{document}